# Robust formation of topological Hall effect in MnGa/heavy metal bilayers

K. K. Meng[1, *], X. P. Zhao[2], P. F. Liu[1], Q. Liu[1], Y. Wu[1], Z. P. Li[1], J. K. Chen[1], J. Miao[1], X. G. Xu[1], J. H. Zhao[2] and Y. Jiang[1, *]

[1] *School of Materials Science and Engineering, University of Science and Technology Beijing, Beijing 100083, China*

[2] *State Key Laboratory of Superlattices and Microstructures, Institute of Semiconductors, Chinese Academy of Sciences, Beijing 100083, China*

**Abstract:** We have investigated the topological Hall effect (THE) in MnGa/Pt and MnGa/Ta bilayers induced by interfacial Dzyaloshinskii-Moriya interaction (DMI). The most evident THE signals have been found based on the MnGa films with small critical DMI energy constant $D_c$. The large topological portion of the Hall signal from the total Hall signal has been extracted in the whole temperature range from 5 to 300 K. These results open up the exploration of the DMI induced magnetic behavior based on the bulk perpendicular magnetic anisotropy materials for fundamental physics and magnetic storage technologies.

*Authors to whom correspondence should be addressed:
*kkmeng@ustb.edu.cn
*yjiang@ustb.edu.cn

The exchange interactions allow the magnetic moments in a solid to communicate with each other and lie at the heart of the phenomenon of long range magnetic order [1]. In the early days of quantum mechanics, the Heisenberg exchange interaction was recognized to determine the types of magnetic ground state. It can be shown as a relatively simple form $J_{12}\mathbf{S_1}\cdot\mathbf{S_2}$, where $J_{12}$ is the exchange constant, $\mathbf{S_1}$ and $\mathbf{S_2}$ are the total spin of two nearby atoms. The sign of $J_{12}$ (positive or negative) determines the coupling modes of $\mathbf{S_1}$ and $\mathbf{S_2}$ (ferromagnetic or antiferromagnetic). However, in reality, the spin-orbit coupling (SOC) will also exert influence on the exchange interactions. On one hand, it connects the magnetization direction to the crystal lattice, and the resulting variation of magnetic energy is referred to as magnetic anisotropy. On the other hand, in the magnetic systems that lack inversion symmetry (whether due to underlying crystal structure or interfaces), SOC can combine with the exchange interaction to generate an anisotropic exchange interaction that favors a chiral arrangement of the magnetization [2, 3]. This is known as the Dzyaloshinskii-Moriya interaction (DMI), which has a form $\mathbf{D_{12}}\cdot(\mathbf{S_1}\times\mathbf{S_2})$. The vector $\mathbf{D}_{12}$ depends on the details of electron wave functions and lies parallel or perpendicular to the line connecting the two spins, depending on the symmetry and the precise crystalline structure. Contrary to the Heisenberg exchange interaction, which leads to collinear alignment of lattice spins, the form of DMI is therefore very often to cant the spins by a small angle. If DMI is sufficiently strong enough to compete with the Heisenberg exchange interaction and the magnetic anisotropy, it can stabilize long-range magnetic bubble domains such as skyrmion. When a conduction electron passes through a magnetic bubble domain, the spin of the conduction electron adiabatically couples to the local spin and experience a fictitious magnetic field (Berry curvature) in real space, which deflects the conduction electrons perpendicular to the current direction. Therefore, it will cause an additional contribution to the observed Hall signals that has been termed topological Hall effect (THE) [4, 5]. THE can be described by the same theoretical scheme as the intrinsic anomalous Hall effect (AHE), which has been clarified to be a Berry phase in momentum space [6].

The recently reported THE in the bulk non-centrosymmetric crystalline structure in B20-type magnets including MnSi, MnGe and FeGe have demonstrated its promising tool for probing DMI [7, 8]. Besides the bulk-type DMI, recent extensive experiments have been focused on the interfacial-type DMI in heavy metal/ferromagnet (HM/FM) bi- and multilayers due to the inherent tunability of magnetic interactions in two dimensions [9-13]. To induce a much larger DMI at the interfaces, Moreau-Luchaire *et al*. have designed the Co-based multilayered thin films in which the Co layer is sandwiched between Ir and Pt layers, which will lead to additive interfacial chiral interactions that increase the effective DMI of the magnetic layer since the two HMs induce interfacial chiral interactions of opposite symmetries and parallel $\mathbf{D}_{12}$ [9]. By harnessing the large and opposite signs of DMI generated from Fe/Ir and Co/Pt interfaces, Soumyanarayanan *et al.* achieved substantial control over the effective DMI governing skyrmion properties in the Ir/Fe/Co/Pt multilayers [12]. The inclusion of Fe results in the gradual formation of a Fe/Ir interface and corresponding suppression of the Co/Ir boundary, leading to increasing DMI and evident THE. More importantly, by varying the ferromagnetic layer composition and thickness, they can modulate the critical DMI energy constant $D_c = 4\sqrt{AK}/\pi$, where $A$ is the exchange constant and $K$ the anisotropy constant. When the effective DMI energy constant $D$ is larger than $D_c$, the skyrmions become thermodynamically stable entities, which is particularly important for technological applications.

In this work, we present the large THE in MnGa/Pt and MnGa/Ta bilayers, in which the MnGa layer is bulk perpendicular magnetic anisotropy (PMA). By varying the growth parameters, we can control $A$ and $K$ of the MnGa layers to decrease $D_c$, and thereby enhance the DMI contribution in the MnGa/HMs bilayers. We clearly demonstrate the extraction of the large topological portion of the Hall signal from the total Hall signal in the whole temperature range from 5 to 300 K and determine the magnitude of fictitious magnetic field.

1-nm-thick MnGa films were firstly grown on 100-nm-thick GaAs buffered semi-insulating GaAs (001) substrates by molecular-beam epitaxy. The growth temperatures have been set at 80, 60 and 40 ℃ for three samples denoted as sample A, B and C. Then the samples were annealed at 300 ℃ for 1 minute. Finally, another nominal 3-nm-thick MnGa films were grown continually at 300 ℃. The thickness of the MnGa films has been controlled by the flux of Mn and Ga atoms referred to our previous works [14, 15]. Figures 1 (a)-(c) show both the out of plane and in plane hysteresis loops of the three samples. The strong difference between the out of plane and in plane curves reveals giant bulk PMA in these films. From the magnetization measurements we can get the following parameters: saturation magnetization $M_s$, the anisotropy field $H_k$ and uniaxial PMA constant $K = M_s H_k / 2$. The exchange constant $A$ ($\sim 2.3 \times 10^{-12}$ Jm$^{-1}$) is roughly considered to be the same for the three samples [16]. Finally, the $D_c$ can be determined and all the parameters have been summarized in TABLE I. Figs. 1(d)-(f) show atomic force microscopy (AFM) images for three samples. It is found that the grains with the size ranged from 30 to 300 nm have been formed in the Sample A, B and C. The surface roughness $R_q$ of the three samples have also been summarized in TABLE I. Therefore, we will focus on the performance of the magnetic and transport properties based on Sample B, in which the smallest $D_c$ has been found. Although the $R_q$ of Sample B is relatively large, the grains shared the same orientation proved by the XRD measurements as shown in the Figure S1 of Ref[17]. Finally, the Pt and Ta layers with different thickness were then deposited on Sample B, a 1.5-nm-thick Al film has also been deposited on Sample B to prevent oxidation but we denote the Sample B/Al (2 nm) bilayers as single MnGa film in following. To determine the DMI at the MnGa/HMs interfaces, we looked into the field dependence of the Hall resistivities in MnGa/Pt (*t*) and MnGa/Ta (*t*) (*t*=2 and 5 nm) bilayers in the whole temperature range from 5 to 300 K, which are compared with the results in a single MnGa film (Sample B). The results are shown in Figs. 2(a)-(e), it is found that the deposited Pt and Ta layers have given rise to anomaly in the Hall resistivities, which shows a bump or dip during the hysteretic measurements. In the presence of a magnetic bubble domain, the total Hall resistivity can usually be expressed as the sum of various contributions [8]:

$$\rho_H = R_0 H + \rho_A + \rho_T, \quad (1)$$

where $R_0$ is the normal Hall coefficient, $\rho_A$ the anomalous Hall resistivity, and $\rho_T$ the topological Hall resistivity.

In Figs. 2(a)-(e), the THE signals clearly coexist with the large background of normal Hall effect and AHE. We have also measured the temperature dependent resistivities of single Al, Pt and Ta films [17]. Considering both the AHE and THE mostly come from the MnGa layer, both the anomalous Hall resistivity, the topological Hall resistivity and the longitudinal resistivity in this paper have been expressed as those of the MnGa layer with the roughly assumption that each film in the MnGa/Pt and MnGa/Ta bilayers acts as a parallel resistance path. To more clearly demonstrate it, we have subtracted the normal Hall term and the temperature dependence of $(\rho_A + \rho_T)$ have been shown in Figs. 2(f)-(j). After the subtractions, we can further discern the peak and hump structure from the MnGa/HM films in the whole temperature range, which can be attributed to the THE term. On the other hand, the combination of different polarity of the normal Hall coefficient in MnGa and Pt layers has changed the sign of $(\rho_A + \rho_T)$ in the MnGa/Pt (5 nm) bilayers as compared with that in other films. The AHE contribution $\rho_A$ can be expressed as [6]:

$$\rho_A = \alpha M \rho_{xx0} + b M \rho_{xx}^2, \quad (2)$$

where $\rho_{xx0}$ is the residual resistivity induced by impurity scattering, $\rho_{xx}$ the longitudinal resistivity, M the magnetization, and *b* the intrinsic anomalous Hall conductivity. The first term is the extrinsic contribution from both the side jump and skew scattering. Therefore, to extract the THE term $\rho_T$, the magnetoresistance (MR) and magnetization in the whole temperature range should be firstly investigated.

The temperature dependences of the longitudinal resistivity $\rho_{xx}$ in all the films show metallic behavior [17]. MR is expressed as MR=$(\rho_{xx}(H) - \rho_{xx}(0))/\rho_{xx}(0)$, and the results of the single MnGa, MnGa/Pt (5 nm) and MnGa/Ta (5 nm) films have been taken as representatives as shown in Fig. 3. The AMR of single MnGa film shows similar behavior with our previous work. More interestingly, dip structures in the AMR curves have been found in the MnGa/Pt (5 nm) and MnGa/Ta (5 nm) bilayers around zero magnetic field. It indicates that the electrons experience a more complicated electromagnetic field due to the presence of non-zero Berry curvature in real space, which becomes stable and plays a more significant role in the transport properties when the applied magnetic field is smaller than the coercivity. Note that the origin of MR is connected with SOC and its influence on *s-d* scattering, and the orbital magnetic moment and the topology of the Fermi surface will be influenced by the strong SOC of the deposited HM layers. Correspondingly, in this case, the electron energy will be modified by the orbital magnetic moment and the electron velocity gains an extra velocity term proportional to the Berry curvature in real space [18]. At the high field limit, i.e., $\omega_c \tau \gg 1$, where $\omega_c$ is the cyclotron frequency and $\tau$ is the relaxation time, the total current in the crossed electric and magnetic fields is the Hall current, suggesting that even in the presence of AHE, the high-field Hall current gives the "real" electron density. It can explain the larger THE signals appearing between zero field and the coercivities. On the contrary, at the low field limit, i.e., $\omega_c \tau \ll 1$, the Berry phase will induce a linear MR as $\sigma_{xx} \approx \sigma_{xx}^{(0)} + \sigma_{xx}^{(1)}$, where $\sigma_{xx}^{(0)}$ is the zero field conductivity, $\sigma_{xx}^{(0)} \propto B_{eff}$, and $B_{eff}$ is the fictitious magnetic field. It may explain the nearly linear MR behaviors around the zero magnetic field, which is much more evident in the MnGa/Ta (5 nm) bilayer due to a stronger fictitious magnetic field.

On the other hand, the unusual MR can also be ascribed to the field-related spin textures which can be visualized in the M-H curves as shown in Fig. 4. Figs. 4(a) and (b) show the temperature dependence of M-H curves in the MnGa/Pt (5 nm) and MnGa/Ta (5 nm) bilayers, and the magnetization reversal can be determined from the inflection points in the M-H curves by adopting the peaks of the first H derivative of magnetization M as shown in Figs. 4(c)-(f). In addition to the magnetization reversal around the coercivities, a sharp peak has also been observed in the MnGa/Ta (5 nm) bilayer around zero magnetic field, indicating its connection with the magnetic bubble domains. Therefore, the dip structures in the MR curves around zero magnetic fields can be ascribed to the modified electrons scattering during the formation and annihilation of magnetic bubble domains. It is noted that this phenomenon becomes weak as decreasing temperature, indicating the gradually weak contribution of DMI.

After accounting for the longitudinal resistivity and magnetization, now we can extract the THE term $\rho_T$. As the topological Hall resistivity should vanish when the ferromagnetic magnetic collinear state is induced, the coefficients in Equation (2) can be determined by describing the AHE resistivity $\rho_T$ in terms of the longitudinal resistivity $\rho_{xx}$ in a high filed region. The extracted $\rho_T$ in all the films have been shown in Figs. 5(a)-(d). It is found that the positions of the peaks are consistent with the inflection in the raw $\rho_H$ data shown in Fig. 2, and the profile has a characteristic hump shape: weak at low field due to the random distribution of magnetic bubble domains, with a peak at higher field due to its proliferation. It is assumed to be a unique signature of THE, and the temperature dependences of the largest $\rho_T$ in all the bilayers have been shown in Figs. 5(e) and (f). Larger signals have been found for the bilayers with thick HM layers, which can be ascribed to their larger coverage on the MnGa grains and more effective SOC. More significantly, THE is evident in the whole temperature range from 5 to 300 K, which is much different from that in B20-type bulk chiral magnets subjecting to low temperature and large magnetic field [7,8]. Since the fictitious magnetic field acts like the classical magnetic field in the same manner as the normal Hall effect, the measured THE signals in the magnetic bubble domains can be written as $\rho_T = PR_0 B_{eff}$, where $P$ is the spin polarization of charge carriers and determined to be 40% for all the films [19], $R_0$ denotes the normal Hall coefficient, and $B_{eff}$ the fictitious magnetic field [12]. Then the temperature dependences of $B_{eff}$ in all the films can be deduced and have been summarized in Figs. 5(g) and (h). The variation tendency is similar with $\rho_T$, and the largest values at 200 K have been found in both MnGa/Pt (5 nm) and MnGa/Ta (5 nm) bilayers, indicating the existence of stronger DMI at this temperature. The $B_{eff}$ values of MnGa/Ta (5 nm) in the whole temperature are all much larger than the bulk-type DMI derived fictitious magnetic field under low temperature and large field [7, 8].

The fundamental reasons that underlie the interfacial DMI have been addressed in the magnetic systems lacking inversion symmetry. However, in our experiment, the lack of inversion symmetry is proved to be not the major reason for the interfacial DMI in the MnGa/Ta and MnGa/Pt films, which is mainly ascribed to the strong SOC. Furthermore, it should be noted that SOC plays different roles in SHE as compared with the interfacial DMI, since there is no obvious connection between the strength and sign of SHE and those of the interfacial chiral DMI. It is suggested by the similar behavior of $\rho_T$ in the MnGa/Pt and MnGa/Ta films, though the spin Hall angles in Pt and Ta are opposite in sign. Recently, using relativistic first-principles calculations, Belabbes *et al*. have shown that the chemical trend of the DMI in 3*d*-5*d* ultrathin films follows Hund's first rule with a tendency similar to their magnetic moments in either the unsupported 3*d* monolayers or 3*d*-5*d* interfaces [20]. The driving force is the 3*d* orbital occupations and their spin-flip mixing processes with the spin-orbit active 5*d* states control directly the sign and magnitude of DMI. A largest absolute DMI value is obtained in Mn/5*d* films, indicating that DMI does depend critically not only on SOC and the lack of the inversion symmetry, but also on the *d* wave function hybridization of the 3*d*-5*d* interface. Therefore, in addition to SOC, the Hund's exchange and crystal-field splitting of *d* orbitals should also be considered in the origin of DMI. Based on this scenario, to some extent, the large THE in the MnGa/Pt and MnGa/Ta films can be partly ascribed to the degree of hybridization between Mn/Pt and Mn/Ta states around the Fermi level. Mn has five filled 3*d* orbitals with all spin up states, where the spin-up (spin-down) channels are totally occupied (unoccupied), so that all the possible transitions between these states will contribute to DMI. It should be noted that the suitable design of MnGa films will decrease the $D_c$ and make the DMI energy more prominent in the total magnetic energies [21, 22]. In our previous works, we have investigated the spin orbit torque (SOT) in the MnGa/HMs films [14, 15], in which the MnGa layer is continuous film ($R_q$=0.4 nm) and the $D_c$ is calculated to be ~1.11e-3 Jm$^{-2}$ [17]. However, we have not found THE in these films though the value of $D_c$ is very close to that in Sample C as shown in TABLE I. It seems that the grain structures will promote the formation of THE, which needs further investigated. On the other hand, if the DMI strength does not depend on the surface properties of MnGa/HMs, the value of *D* would be around 1.08e-3 Jm$^{-2}$ since a weak THE emerged in Sample C/Pt (5 nm) bilayers [17]. We have also investigated the SOT in MnGa/Pt (5 nm) bilayers based on the three samples. The SOT only happened in the Sample C/Pt bilayers [17], but the magnetization has not been fully switched. The SOT driven magnetization switching is supposed to be subdued for grain films, though the nominal domain wall width $\Delta=\sqrt{A/K}$ is almost the same for Sample B and C, the domain wall motion will be strongly influenced by the grain structures. Furthermore, we roughly determined the same value of *A* for the three samples, which actually will change due to the structural disorders of the MnGa films [23]. Based on a simple assumption in the Stoner model for itinerant ferromagnetism, the smaller saturation magnetization in Sample B and C can be an index of the weak exchange interaction [24]. Totally speaking, the role of DMI in SOT driven magnetization switching and THE based on MnGa films should be further exploited.

In summary, we have demonstrated the presence of large THE in the MnGa/Pt and MnGa/Ta bilayers induced by strong interfacial DMI between HMs (Pt and Ta) and Mn atoms. The largest THE signals have been found based on the MnGa films with small $D_c$, in which the DMI will play a more dominant role in the total magnetic energies. The large topological portion of the Hall signal from the total Hall signal has been extracted in the whole temperature range from 5 to 300 K. These results open up the exploration of the DMI induced magnetic behavior based on the bulk PMA materials for fundamental physics and magnetic storage technologies.

**Acknowledgements:** This work was partially supported by the National Basic Research Program of China (2015CB921502), the National Science Foundation of China (Grant Nos. 51731003,

61404125, 51471029, 51671019, 11574027, 51501007, 51602022, 61674013, 51602025), and the Fundamental Research Funds for the Central Universities (FRF-GF-17-B6).## References

[1] Y. Taguchi, Y. Oohara, H. Yoshizawa, N. Nagaosa, and Y. Tokura, Science **291**, 2573 (2001).

[2] I. Dzyaloshinsky, J. Phys. Chem. Solids **4**, 241 (1958).

[3] T. Moriya, Phys. Rev. **120**, 91 (1960).

[4] J. Ye, Y. B. Kim, A. J. Millis, B. I. Shraiman, P. Majumdar, and Z. Tešanovic, Phys. Rev. Lett. **83**, 3737 (1999).

[5] P. Bruno, V. K. Dugaev and M. Taillefumier, Phys. Rev. Lett. **93**, 096806 (2004).

[6] N. Nagaosa, J. Sinova, S.Onoda, A. HmacDonald and N. P. Ong, Rev. Mod. Phys. **82**, 1539-1592 (2010).

[7] Y. Li, N. Kanazawa, X. Z. Yu, A. Tsukazaki, M. Kawasaki, M. Ichikawa, X. F. Jin, F. Kagawa, and Y. Tokura, Phys. Rev. Lett. **110**, 117202 (2013).

[8] W. Wang, Y. Zhang, G. Xu, L. Peng, B. Ding, Y. Wang, Z. Hou, X. Zhang, X. Li, E. Liu, S. Wang, J. Cai, F. Wang, J. Li, F. Hu, G. Wu, B. Shen, and X.-X. Zhang, Adv. Mater. **28**, 6887 (2016).

[9] C. Moreau-Luchaire, C. Moutafis, N. Reyren, J. Sampaio, C. A. F. Vaz, N. Van Horne, K. Bouzehouane, K. Garcia, C. Deranlot, P. Warnicke, P. Wohlhüter, J.-M. George, M. Weigand, J. Raabe, V. Cros, and A. Fert, Nat. Nanotechnol. **11**, 444 (2016).

[10] S. Woo, K. Litzius, B. Krüger, M.-Y. Im, L. Caretta, K. Richter, M. Mann, A. Krone, R. Reeve, M. Weigand, P. Agrawal, P. Fischer, M. Kläui, and G. S. D. Beach, Nat. Mater. **15**, 501 (2016).

[11] O. Boulle, J. Vogel, H. Yang, S. Pizzini, D. de Souza Chaves, A. Locatelli, T. Onur Menteş, A. Sala, L. D. Buda-Prejbeanu, O. Klein, M. Belmeguenai, Y. Roussigné, A. Stashkevich, S. Mourad Chérif, L. Aballe, M. Foerster, M. Chshiev, S. Auffret, I. Mihai Miron, and G. Gaudin, Nat. Nanotechnol. **11**, 449 (2016).8

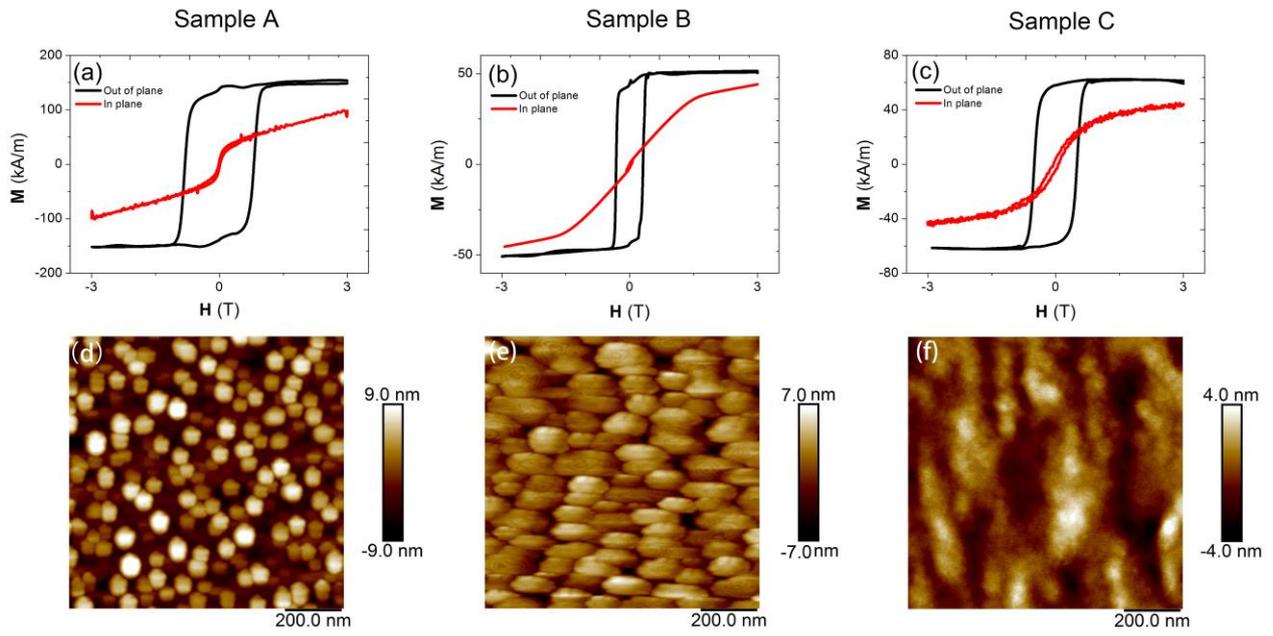

Figure 1. (a)-(c) Hysteresis loops for the out of plane and in plane magnetizations of the three samples at 300 K. (d)-(f) AFM images of Sample A, B and C (scale bar, 200 nm).

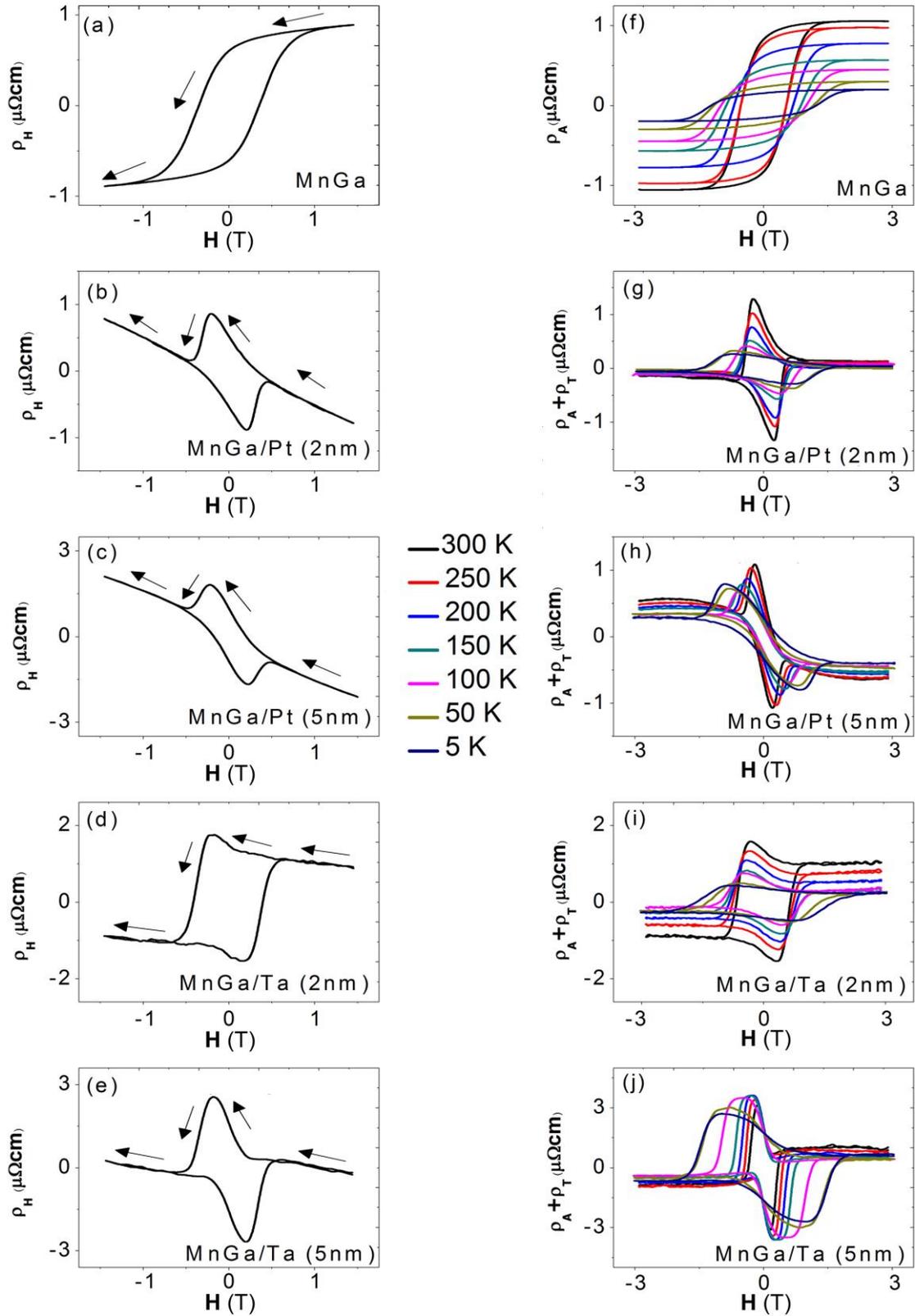

Figure 2. (a)-(e) Total Hall resistivities in the single MnGa, MnGa/Pt (t) and MnGa/Ta (t) (t=2, 5 nm) films at 300 K. The arrows denote the change tendency of hall resistivity when the magnetic field is applied from positive to negative. (f)-(j) ($\rho_A + \rho_T$) *vs* H for the single MnGa, MnGa/Pt (t) and MnGa/Ta (t) (t=2, 5nm) films in the temperature range from 5 to 300 K.

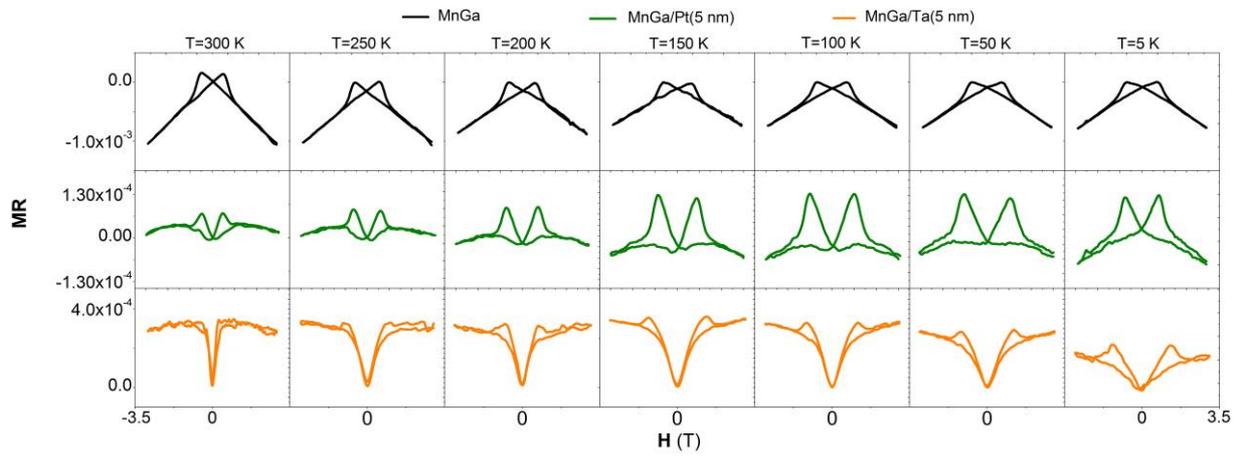

Figure 3. Magnetoresistance of the single MnGa, MnGa/Pt(5 nm) and MnGa/Ta(5 nm) films in the temperature range from 5 to 300 K.

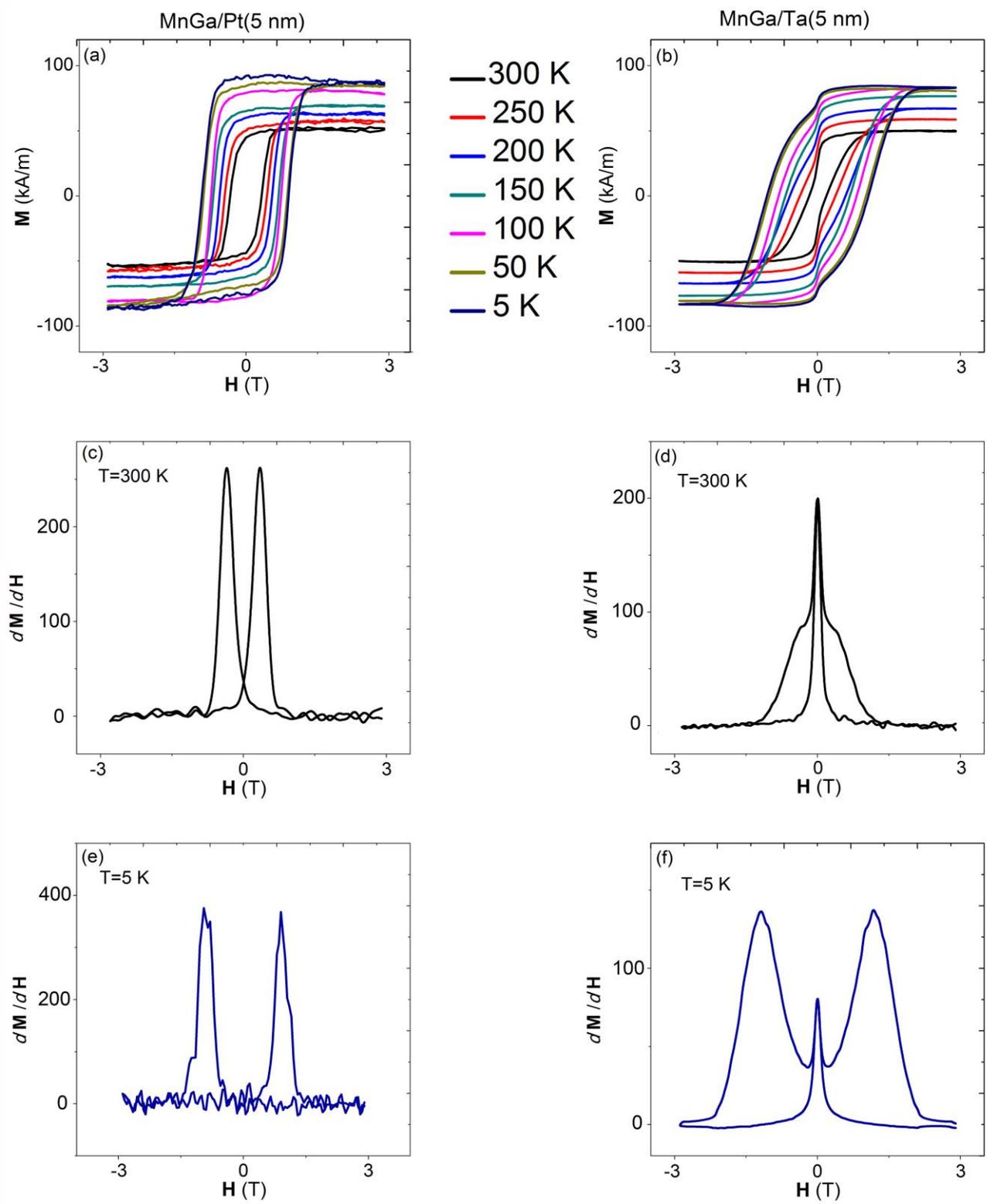

Figure 4. (a) and (b) show the out of plane hysteresis loops for MnGa/Pt(5 nm) and MnGa/Ta(5 nm) in the temperature range from 5 to 300 K respectively. (c) and (d) show the first H derivative of the magnetization M in the two bilayers at 300 K. (e) and (f) show the first H derivative of the magnetization M in the two bilayers at 5 K.

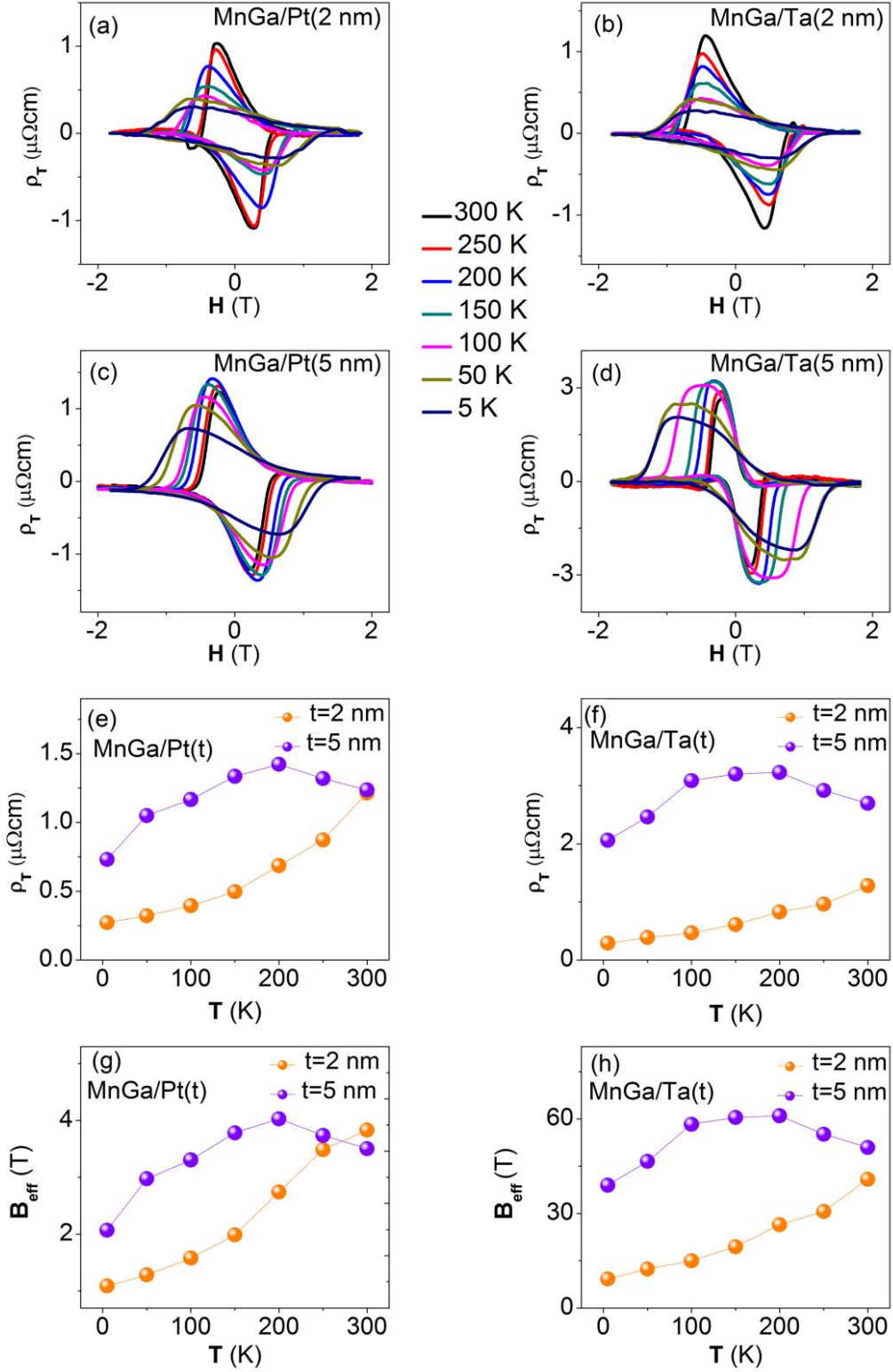

Figure 5. (a)-(d) $\rho_T$ (H) in MnGa/Pt(t) and MnGa/Ta(t) (t=2, 5 nm) films at different temperatures. (e) and (f) Temperature dependence of $\rho_T$ for all the films. (g) and (h) show the temperature dependence of fictitious magnetic field $B_{eff}$ in all the films.

TABLE I. Saturation magnetization $M_s$, the anisotropy field $H_k$, uniaxial PMA constant $K$, the critical DMI energy constant $D_c$ and the surface roughness $R_q$ of the three samples.

| Sample | $M_s$ (kAm$^{-1}$) | $H_k$ (T) | $K$ (Jm$^{-3}$) | $D_c$ (Jm$^{-2}$) | $R_q$ (nm) |
|---|---|---|---|---|---|
| A | 145 | 13.7 | 9.93e5 | 1.94e-3 | 3.6 |
| B | 51 | 4.5 | 1.15e5 | 6.52e-4 | 2.6 |
| C | 63 | 9.5 | 3.14e5 | 1.08e-3 | 1.1 |